\documentclass[12pt,letterpaper]{article}
\usepackage{amsmath,amssymb,pgf,pgfarrows,pgfnodes,float,appendix, hyperref}
\usepackage{graphicx}
\usepackage{subfigure}
\usepackage[margin=0.9in]{geometry}

\newcommand{\be}{\begin{equation}}
\newcommand{\ee}{\end{equation}}
\newcommand{\bea}{\begin{eqnarray}}
\newcommand{\eea}{\end{eqnarray}}

\title{{\rm\footnotesize \qquad \qquad \qquad \qquad \qquad \ \qquad \qquad \qquad \ \ \ \ \ \                  RUNHETC-2016-21,}\vskip.5in    Note on a Paper by Ooguri and Vafa}
\author{Tom Banks\\
Department of Physics and NHETC\\
Rutgers University, Piscataway, NJ 08854\\
E-mail: \href{mailto:tibanks@ucsc.edu}{tibanks@ucsc.edu}}

\date{}
\begin{document}
\maketitle

\begin{abstract}
In a recent paper, Ooguri and Vafa\cite{ovkf} argued that a mild extension of the Weak Gravity conjecture\cite{weakgrav} led to the conclusion that the only models of quantum gravity in AdS space with ``radius large compared to the string scale" were models with exact AdS-SUSY.  This note clarifies certain obscure parts of their argument by reinterpreting it as a statement about brane configurations in flat space. In that context the statement is that stable non-supersymmetric configurations of branes are characterized by charges in the torsion subgroup of K-theory and their only long range fields are gravitational and scalar.  The field equations do not have large radius near horizon AdS solutions. Instead the horizons are Schwarzschild black branes.  This leads us to conjecture that for $d \leq 11$ the K theory charges are bounded by a small integer, and there can be no large gravitational back reaction in the near horizon limit, for stable configurations. Unbounded values of these discrete charges would violate the Covariant Entropy Bound. We discuss a counterexample to the conjecture in $AdS_3$. We find similar counterexamples in higher dimensional AdS spaces, but argue that none of them lead to SUSY violating models of quantum gravity in Minkowski space.
\end{abstract}

\section{Introduction}

A large radius AdS/CFT model of quantum gravity is one for which there is a gravitational dual whose low lying spectrum consists of only the modes of a finite number of fields on a manifold $AdS_d \times {\cal K}_D$ with radius of curvature of both manifolds satisfying $R \gg L_S \gg L_P$.  $L_S$ is the scale defined by the string tension in the weak coupling regions of moduli space . All other states should be parametrically heavier, meaning that they correspond to conformal primaries with dimensions scaling like a positive power of $R/L_S$ .  The AdS black hole entropy formula tells us that $R/L_P$ is large when the coefficient $c_F$ of the free energy of the CFT in flat space, is large.  However, even in CFTs with large $c_F$ the degeneracies of dimensions grow like an exponential of a power of the dimension.   Large radius duals occur only when there is a gap in the dimension spectrum, where the degeneracy only grows as a power.  If $C_F$ is large, but there is no gap then, by abuse of language, we say that the AdS radius is of order the string scale\footnote{I first proposed this characterization of large radius during the ITP workshop in Santa Barbara in 1998. It did not become popular until people figured out how to do useful things with it\cite{useful}. }.  

For a long time\cite{previous}, I've conjectured that the only CFTs with such a parametrically large gap in the spectrum were Superconformal theories.  The primary motivation for this was my previous conjecture that models of quantum gravity in Minkowski space had to be exactly Super-Poincare invariant.  If one has a sequence of models with arbitrarily large $R/L_P$ in which SUSY is violated by an amount independent of $ R/L_P$ then one might argue that one gets a SUSY violating theory in Minkowski space in the limit $R/L_P \rightarrow \infty$. We'll see below that there are counterexamples, in which the AdS radius can be large without SUSY, but SUSY violation goes away in the infinite radius limit.

Note that perturbing a large radius SCFT by a SUSY violating relevant operator does not violate this conjecture, as long as the RG flow does not approach a large radius non-SUSic CFT in the infrared.  In such models, the SUSY violation goes away in the infinite radius limit.  One piece of evidence for the conjecture is that many SUSY violating holographic RG flows, which do appear to approach another large radius AdS space, have Breitenlohner-Freedman forbidden tachyons in their spectrum of fluctuations.  This is in sharp contrast to flows which partially break SUSY, and lead to new stable large radius fixed points. T. Dumitrescu and N.Bobev have pointed out that there may be some counterexamples to this phenomenological observation\cite{tachfreeholorg}.    We'll see below that there is evidence for non-SUSic CFTs, with large radius, where the SUSY violation vanishes in the large radius limit, because the shifts in dimension from a supersymmetric theory are order $1$ in that limit.

Finally, there have been concerted efforts to find large radius CFTs without SUSY, by a number of people. My own work on this topic has remained unpublished because I was unsuccessful in finding examples but unable to prove that there were none. I suspect that other attempts to do this met a similar fate. There is a recent no go theorem for a certain class of CFTs in $d = 2$\cite{findthegap}, but there is also a two dimensional counterexample, which we will discuss below.

The authors of \cite{ovkf} provide a dynamical explanation of why there are no  stable large radius SUSY violating models in AdS space.  They observe that known stable models are supported by the flux of some $p$-form field.  A slight strengthening of the weak gravity conjecture implies that there is a brane with small enough tension to nucleate and reduce the flux by one unit, following the argument of \cite{bunster} .  

This argument, while compelling, has some obscure features.  One argues for the nonexistence of a CFT by showing that its putative gravity dual is ``unstable".  What could a CFT decay to? On the gravity side, AdS is like a closed box.  Massive states cannot reach the boundary, massless particles bounce off it. The Coleman-de Luccia instantons for decay of AdS space always lead to singularities inside the expanding bubble of ``stable vacuum".  Investigation\cite{brhsm} of this question has led to the conclusion that the only such gravitational ``instabilities" that make sense correspond to putting a relevant perturbation of the CFT on a boundary de Sitter space, rather than to an instability.

However, if one modifies the conjecture just a little bit, all of these questions are resolved.  All of the known large radius AdS models have been derived by starting with a configuration of branes in flat space, or near the tip of a non-compact geometry which is the product of flat space coincident with the brane world volume, and a cone over some compact manifold.  There are other models\cite{wrappedbranes} in which branes are wrapped on cycles of a non-compact Calabi-Yau or $G_2$ manifold.

There is a large literature on the study of decays of non-supersymmetric brane configurations\cite{nonsusybranes}.   In the non-compact geometries in question, radiation can go off to infinity, so the notion of decay is perfectly well defined.  The stable remnants can be large or small collections of supersymmetric branes, or stable non-supersymmetric branes.  The latter are classified by the torsion subgroup of K theory charges\cite{ktheory} or their M-theory generalization\cite{dmw}. From now on we'll use the term K-theory charges to refer to both of these cases. What is relevant for classifying stable non-BPS configurations are the discrete torsion subgroups of the K-theory group. 
Co-dimension $D$ branes in flat space can be thought of as sitting at the tip of a cone over a sphere in the transverse dimensions. By replacing the sphere by some other manifold, one can obtain models, following\cite{klebwit} with reduced supersymmetry.   In models with branes wrapped on cycles of or at the tips of non-compact SUSY preserving manifolds one can also imagine constructing SUSY violating brane configurations classified by torsion subgroups of the K-theory group.  Let us imagine that we can find sequences of examples where the order of the torsion subgroup is unbounded.

Of course, one might try to argue there is an analog of the \cite{ovkf} tunneling instability for these configurations as well.  It's much harder to make the argument for these hypothetical configurations.  In the case of branes coupled to a long range non-gravitational field, the balance of the long distance gauge and gravitational forces between a brane configuration with large charge and one where the charge is separated into a lower charge cluster and a single well separated brane of small charge, depends on a single parameter, the charge to tension ratio.  The extended weak gravity conjecture implies that the repulsive gauge force always wins, so that the energetics favors a tunneling instability.   

If there are sequences of models with unbounded torsion K theory charge, then the long range force is purely gravitational, so there are surely stable bound configurations with energy density concentrated on microscopic scales.  However, the same argument suggests that these regions would not have near horizon regions with the geometry of AdS and large curvature radius, even though the brane tension might be unbounded.  If there is an large radius AdS space, it should be a solution of the low energy effective field theory in the near horizon region.  That effective field theory is simply supergravity, with vanishing c.c. (because the branes are embedded in a higher dimensional flat space), and no source for any of the non-gravitational long range fields, except perhaps scalars.  There are no AdS solutions to the field equations.  Generically, if the tension is large, it will be a Schwarzschild or Schwarzschild-dilaton black brane, which will decay to a string scale size by Hawking radiation.

This argument points up the fact that the existence of models with unbounded torsional K-theory charge would violate (the brane analog of) the covariant entropy bound.  That is, we could find brane configurations of unbounded entropy density, whose horizon area was microscopic.  This leads us to conjecture that no such infinite sequences of models exists. Let's briefly review the argument for this claim\cite{tbjs}:  Suppose we had a sequence of SUSY preserving solutions of the field equations of SUGRA, of the form $M^{1,d} \times {\cal K}$ where ${\cal K}$ is a SUSY preserving manifold of dimension $9 - d$ in string theory, or $10 - d$ in M-theory models.  ${\cal K}$ is a cone over a sequence of compact manifolds with unbounded torsion $K$-theory charge, or the product of a compact manifold with such a cone.  Now consider the SUGRA fields generated by putting a large number $N$ of branes plus anti-branes on this geometry, with very large torsional $K$-theory charge.  The branes may lie at the tip of the cone or be wrapped on cycles. This gives a black brane of large tension, whose only long range fields are gravitational and scalar fields.

The quantization of perturbative modes around this configuration leads to Hawking radiation and the black brane loses tension in a thermal process.
Unlike the case of branes with long range $p$-form gauge fields, there is no preference for emission of net torsional $K$-theory charge, to leading order in the semi-classical expansion.  The configuration evolves to one of order Planck or string scale tension, which still retains a large charge\footnote{Strictly speaking, if $d \geq 1$ then we should be using the phrase charge per unit volume, rather than charge.  All the charges of extended objects are infinite.}. At some critical finite value of the order of the torsion subgroup of the $K$-theory group, one would violate the brane analog of the covariant entropy bound.  

Should there in fact be a brane analog of the CEB?  In flat space, where the manifold ${\cal K}$ is a sphere, we simply wrap the branes on a torus and get particles propagating in $10 - d$ or $ 11 - d$ dimensional space-time.  As long as $10 - d \geq 4$ we would get microsystems violating the ordinary CEB.   This is not as clear for particles propagating on a $CY_d$ or $G_2$ cone.  One would have to establish the existence of Schwarzschild black holes on such space-times and redo the Hawking calculation.  It is hard to see how this could fail for black hole radii much smaller than the radii of curvature of the cone.  We're thinking about black holes whose horizon shields the singularity at the tip of the cone.   It would seem that the gravitational duals of finite temperature states of models like the Klebanov Witten model on a three torus, should give us examples of such black hole geometries. In \cite{bena} black holes in Klebanov-Tseytlin and Klebanov-Strassler space-times were constructed, so it seems likely that the generalized covariant entropy bound for black branes is valid.

Thus, there is a very sharp mathematical conjecture, which is motivated by this argument.  There should not exist sequences of smooth compact $CY_3$ or $G_2$ manifolds with unbounded order of the torsion subgroups of their K-theory groups. The same should be true for cones over compact manifolds.  Note that this conjecture is true for $CY_2$ manifolds.  G. Moore points out that the results of \cite{achdoug} might be a good starting point for work on this conjecture.

\subsection{A Counterexample That Isn't}

Another approach to finding counterexamples to the conjecture of \cite{ovkf} also involves stable non-supersymmetric configurations of D-branes.  Consider such a configuration, $B_K$ with non-vanishing torsional K-theory charge, formed by merging equal stacks of branes and anti-branes.  Now consider a collection of a large number of BPS branes, whose near horizon limit is a large radius AdS space, separated from $B_K$ by a large transverse distance. The only long distance force between the two branes is gravitational, so there will be a bound state, which violates SUSY.  The gravitational back reaction of the low tension configuration $B_K$ will be small, so the near horizon region will still be a large radius AdS space.  

However, as pointed out by G. Moore\footnote{G. Moore, private communication.}, the bound state will have the a non-zero length scale: the expectation value of the distance between the BPS stack and the brane $B_K$.  Thus, it is properly viewed as a SUSY violating relevant perturbation of the original CFT.  General quantum field theory wisdom assures us that such models exist.   Their gravity dual is a domain wall space-time, which is possibly singular in the region corresponding to the infrared.  The singularity corresponds to the massive infrared physics of the perturbed theory, and cannot usually be studied by purely geometrical means.

\section{A Counter-example in $AdS_3$ and Some Higher Dimensional Generalizations}

There is of course a longstanding counter-example\cite{evaetal} to the conjecture that there cannot by SUSY violating CFTs dual to large radius $AdS_3$ geometries.  Recent work\cite{dongetal} has elucidated the way in which the SUSY breaking mechanism works in the bulk.  It is simplest to describe these models in the boundary field theory.  One starts with one of the familiar sequences of large radius models with $(2,2)$ superconformal symmetry.   All such models contain holomorphic and anti-holomorphic $U(1)$ currents $J_{L,R}$.  The stress tensor is the sum of Sugawara stress tensors $T_{L,R}$ for these two currents, and a commuting singlet stress tensor $T_S$. The connected $n > 2$ point functions of the currents vanish.

All such models have an exactly marginal perturbation $\delta {\cal L} = g J_L J_R$ .  This shifts the dimensions of operators in a way that depends on their R charge, and breaks supersymmetry.  To make the SUSY breaking large in inverse AdS  radius units, one must go to the limits of the moduli space of $g$, which are infinite distance away in the Zamolodchikov metric\footnote{I thank Xi Dong for a reminder of this.}.  This corresponds to a decompactification limit.  The generators $J_{L,R}$ are part of the $SU(2)_L \times SU(2)_R$ isometry group of the sphere $S^3$.  The coupling $g$ breaks this symmetry and corresponds to a deformation of the spherical geometry.  At the boundaries of moduli space, one dimension of the sphere becomes infinitely larger than the other two.  The latter remain at the AdS radius $R$.  

If one takes $R/L_P \rightarrow \infty$ at the supersymmetric point, one obtains a theory in six dimensional Minkowski space, with a finite sized $T^4$ or $K_3$ Kaluza-Klein manifold.  The same holds true if we take $R/L_P \rightarrow \infty$ at some fixed point in $g$ moduli space.  SUSY breaking vanishes in the flat space limit.  On the other hand, taking $g$ to the boundaries of moduli space for fixed $R/L_P$, one obtains a logarithmic CFT, with a continuous spectrum of operator dimensions, and it's not clear what the space-time interpretation of the correlation functions are\footnote{One can ask a similar question if one takes the $T^4$ radius to infinity at fixed $R/L_P$ and its not clear what the answer is in that case either.}.  Finally, one can imagine taking a correlated limit in which $R/L_P \rightarrow \infty$, while $g$ goes to the boundary of its moduli space, in such a way that the splittings in super-multiplets remain finite.  Again, it's not clear what the interpretation of that limit is, but it clearly violates some of the rotational symmetries of $6$ dimensional Minkowski space.

One can obtain the coupling $g J_L J_R$ by coupling the original model to a $U(1)$ gauge field via
\begin{equation} {\cal L} = \frac{1}{2 g} (A_{\mu} - \partial_{\mu} \theta) )^2 + A_{\mu} J^{\mu} , \end{equation} where $J^{\mu}$ is the anomaly free vector combination of $J_L + J_R$ . Here $\theta$ is a periodic Stueckelberg field, which can be gauged away apart from vortex configurations.  In higher dimensional CFTs, one can imagine doing something similar.  For example, one could gauge the $SO(8)$ R-symmetry of the $AdS_4 \times S^7$ CFT, or some subgroup of it, and add a level $k$ Chern-Simons term for the gauge field.  This appears to break SUSY while preserving conformal invariance at the classical level of the bulk theory, because there are no couplings that can run.  
However, one must examine carefully the possibility that this perturbation induces running of triple trace marginal couplings.  

There are operators in the model, which can be thought of as originating as BPS trilinears of the scalar fields of maximally SUSic Yang Mills theory.  Near the $AdS_4 \times S^7$ fixed point, they give rise to primary fields in short representations of the superconformal algebra, which transform as a combination of singlet and traceless symmetric rank two tensor of $SO(8)$.  There are three invariant triple trace marginal operators, which one can construct from these fields.  In leading order in large $N$, they are all marginal operators and their correlation functions can be calculated by a change of boundary conditions of the bulk SUGRA fields.  At order $1/N^2$, their dimensions are shifted.  In the analogous case of double trace operators in $4$ dimensional ${\cal N} = 4$ SYM theory, the shift makes the operators marginally relevant\cite{rastelli} .  In leading order in large $N$ the Chern-Simons perturbation of the model does not change the boundary conditions of bulk scalars and does not induce these triple trace couplings.  It's likely that there will be contributions to the running of these couplings from the Chern-Simons interaction at higher orders in the $1/N$ expansion.  Then the induced couplings will give even higher order contributions to their own running since there are non-vanishing three point functions among the triple trace operators.  On their own, the effect of these terms is to make the couplings marginally irrelevant, but they are smaller than contributions to the dimensions analogous to those calculated by \cite{rastelli}.  If those are marginally relevant, we would break conformal invariance.  Note however, that by adding an explicit triple trace coupling at leading order in large $N$ in addition to the Chern-Simons coupling, we could tune away this instability and find a controllable large $N$ fixed point, which violated SUSY.  These statements need to be checked more carefully, but seem like the basis for a proof of the existence of such models.

The shifts in dimensions between the SUSY violating and SUSY preserving fixed points would, for finite $k$ be order $1$ multiples of the inverse AdS radius. Thus, they will not affect the existence of a large gap where the multiplicity of operators of dimension $\Delta$ grows only like a power of $\Delta$.  We therefore have a robust construction of a non-supersymmetric CFT with large radius dual.  One cannot blame this on a decoupled sector.   Note however that the shifts of dimensions of singlet operators go to zero in the large $k$ limit, which is the only extreme regime in the perturbed model parameter space.   Non-singlet operators are projected out in this limit. Thus, one cannot construct a model of this type that leads to a SUSY violating theory of gravity in Minkowski space.  This seems to be a counterexample to the conjecture of \cite{ovkf}, which cannot be dismissed by saying that it occurs in a decoupled sector.  Similar deformations of ABJM theories, with similar properties, would lead to the same conclusion.   In the case where the group that is gauged is a product of $U(1)$'s, the Chern-Simons coupling can be replaced by the $B_i d A_i$ couplings of \cite{ksw} .  Note however that those authors did not couple gauge fields to R symmetry currents.

The $SU(4)$ R symmetry of maximally super-symmetric Yang Mills is anomalous, but it has a variety of anomaly free subgroups.  There are two linear combinations of the Cartan generators, which are anomaly free.  One can think of them as the Cartan generators of the $SO(4)$ subgroup, under which the real and imaginary parts of the supercharges transform as a pair of four vectors. We can couple two $U(1)$ gauge potentials $A_i$, with field strengths $F_i$ to these conserved currents. The topological Lagrangian
\begin{equation} {\cal L}_{top} = k_{ij}  B_i  F_j , \end{equation} where $B_i$  are two form gauge potentials, makes sense only when the coefficients $k_{ij}$ are appropriately quantized.  Thus, since the new couplings cannot run, we get a new CFT, which breaks all of the SUSY, if we can control the running of other marginal couplings.  One would imagine that a similar strategy works for higher dimensional CFTs, where in $d$ (boundary) space-time dimensions $B_i$ are replaced by $d - 2$ forms\footnote{Here $B_i$ and $F_i$ are labeled by the Cartan generators of an anomaly free subgroup of a continuous R symmetry group.}.   If the resulting models are conformally invariant, their properties will be analogous to the $AdS_4$ models with Chern-Simons terms.  SUSY breaking will vanish in the limit of large AdS radius.  These topological perturbations of higher dimensional CFTs have not been studied in the literature, so we cannot be sure that dimensions really shift, but the argument that the large $k$ limit projects out the non-singlets would seem to go through in this case as well, so that SUSY is broken.

For $5$ and $6$ dimensional SCFTs, there do not seem to be any ``multiple trace" marginal couplings, which could interfere with this argument.  However, the four dimensional case is more problematic because the large radius condition is obeyed only at large 't Hooft coupling.  As in the SUSY violating ``orbifold" theories of \cite{orbifold}, the real problem is that at order $1/N^2$ one expects the 't Hooft coupling to run according to
\begin{equation} \dot{\lambda}_{tH} = \frac{\beta (\lambda_{tH}) }{N^2} . \end{equation}
There is no line of fixed points, and no reason for any fixed point at large $\lambda_{tH}$.   At weak coupling the absence of a line of fixed points was shown in\cite{dymkleb}, so if these theories have fixed points at all, then even at large $N$ the AdS radius will be of order string scale.

\section{Conclusions}

We have reformulated the conjecture of \cite{ovkf} as a statement about non-supersymmetric configurations of branes in flat space or at the tips of cones ${\cal K}$ whose geometry admits Killing spinors.  These can be classified by torsion K-theory charges or their M-theory analog.  The long distance fields of such configurations are purely gravitational/scalar and they do not have AdS near horizon regions, even if it were possible to have unbounded torsional K-theory charge.  We suggest that unbounded torsional K theory charge is impossible in space-times that are products of Minkowski space and a supersymmetric cone.  This would lead to a violation of a brany analog of the Covariant Entropy bound.  For branes in flat space this reduces to the ordinary covariant entropy bound by compactifying the brane world volume on a spatial torus.  For general supersymmetric cones, one would have to establish the existence of black hole configurations on the space-time $R \times {\cal K}$ to make a similar argument.  This is certainly plausible when the black hole radius is small compared to the radii of curvature of the cone.   For example, a finite temperature state of the Klebanov Witten CFT and its generalizations, compactified on a three torus, should give us black holes on a class of Calabi-Yau cones. The mathematical question of the spectrum of torsional K theory charges over such spaces can be studied independently of such hand waving arguments.

We also studied two classes of counterexamples to the conjecture of \cite{ovkf} : one achieved by coupling a gauge field to an R symmetry current, and the other by dropping stable non-supersymmetric branes into large stacks of BPS branes with a large radius AdS near horizon region.  We argued that the second class violated conformal invariance and reduced to the well known procedure of perturbing a large radius supersymmetric model by a SUSY violating relevant operator.

The two dimensional version of the gauged R symmetry models was dismissed by Ooguri and Vafa as special to the decoupled ``topological sector" of supersymmetric $AdS_3$ models.  We were able to find similar models in higher dimensions, which do not succumb to that criticism.  They have discrete rather than continuous parameters.  In all cases we argued that these models did not lead to supersymmetry violation in Minkowski space when the large radius limit is taken.  The original conjectures of\cite{previous} were based on the assumption that sequences of large radius SUSY violating CFTs would lead to models that violated SUSY in flat space.  There is, in the opinion of the present author, compelling evidence that no such models exist.
What we've seen instead is that, like large radius SCFTs perturbed by a SUSY violating relevant operator, the SUSY violation in the counterexamples to the conjecture of \cite{ovkf} vanishes in the flat space limit.  It should be emphasized that in $d \geq 4$ the SUSY violating CFTs apparently obtained by perturbing a SCFT by a topological $B_{d - 2} F$ term have yet to be investigated. We have presented plausible arguments that they exist for $d > 4$.  Four dimensional examples of large radius SCFTs depend on lines of fixed points, which are very sensitive to small perturbations.

\vskip.3in
\begin{center}
{\bf Acknowledgments }\vskip.2in \end{center}

The work of T.Banks is {\bf\it NOT} supported by the Department of Energy. I would like to thank  N. Seiberg, C. Vafa, H. Ooguri, Scott Thomas, D. Shih, L. Rastelli, D. Freedman, G. Horowitz, T. Dumitrescu, N. Bobev, I. Klebanov and especially Z. Komargodski, Xi Dong and G. Moore, for crucial input and comments.

\section{Appendix: Stable dS Space}

The second reference in \cite{ovkf} attempted to extend the arguments about instability of large radius SUSY violating AdS space to the case of positive c.c. .  No such extension can be justified.  We have two well established mathematically consistent frameworks for studying the problem of spacetimes with negative c.c. . The first is the general theory of CFT and the second is the construction of AdS regions near the horizon of collections of branes in string/M-theory on non-compact SUSY preserving space-times.  By contrast here is no agreed upon mathematical theory of dS space, or of the ``landscape of meta-stable dS vacua" .  The {\it only} extant mathematically well defined model of dS space is the conjecture of Fischler and the present author that the quantum theory of space-times that asymptote to stable dS space in the future, can be formulated in a finite dimensional Hilbert space. It may be incorrect, but can only be shown to be so by exhibiting an alternative mathematical definition with superior properties.

The only reliable information we have about what the quantum theory of dS space might look like comes from Coleman de Luccia instantons\cite{cdl} and black hole entropy formulae .  For tunneling between two dS minima, we find a law of detailed balance, which is consistent with the idea of finite dimensional Hilbert space, and suggests an interpretation of localized energy in dS space as a parametrization of the number of constraints on the state of the degrees of freedom responsible for the entropy of the dS vacuum ensemble.  That is, localized energy corresponds to freezing of order $E/T_{dS}$ of the fundamental q-bits.  This interpretation of energy explains the dS black hole entropy formula, and is also useful for understanding the production and decay of black holes in Minkowski scattering processes\cite{holonewton2}.

For transitions from dS minima into negative c.c. Big Crunch regions CDL instantons divide up into two classes, distinguished by whether or not the Minkowski minimum obtained by adding a constant negative potential to cancel the dS c.c., has a positive energy theorem or not\cite{abj}.  For those cases that do have a positive energy theorem in the limit of canceled c.c.  the tunneling probabilities are entirely consistent with the idea that the Hilbert space is finite dimensional and that the singular spacetime to which one tunnels is a meta-stable configuration of much lower entropy than the dS vacuum.  This is also consistent with the observation that the maximal causal diamond in the crunching region has area much smaller than the dS horizon area.

By contrast, the situation with an unstable flat space limit, has no obvious quantum interpretation.  Large radius dS space and Minkowski space have high entropy black hole excitations, which decay by Hawking radiation.  Even in flat space, the expanding bubble of crunching spacetime does not swallow up the space inside the black hole horizons.  Thus, the endpoint of the hypothetical decay of flat space depends on which excited state of the ``false vacuum" one is studying.  The different endpoints are all singular space-times with multiple singular asymptotic regions, each of different entropy.  The black holes cannot decay by emitting particles into the singular spacetime which has swallowed up the region outside their horizons.  This does not at all resemble the decay of a false vacuum in quantum field theory, and no one has proposed a plausible quantum mechanical interpretation of such a system.

Firm theoretical evidence in favor of one or another model of the quantum theory of dS space is likely to elude us for some time.  Much of the sentiment in favor of the instability suggested by Freivogel and Kleban seems to stem from the paper of \cite{DKS} .  One should first of all point out that the problem with stable dS space described in that paper is really a phenomenological rather than a mathematical problem, and that it is only an unavoidable problem if one makes the assumption that our own universe began as a low entropy fluctuation of a finite ensemble, whose time development is determined by a time independent Hamiltonian.  This assumption was first entertained by Boltzmann, and shot down by his assistant Schutz a short time later.  Since the space-time of our observed cosmology has no time like Killing vector, there is no reason to make this assumption.  For a finite system with time dependent Hamiltonian, which asymptotes to the Hamiltonian of dS space, one can \cite{holoinflation2} make a model of what we observe before the dS era is reached.   In that model, the answer to Boltzmann's question of why the universe began with low entropy is that states with localized energy must have low entropy, because of the fundamental definition of energy as a parametrization of the number of constraints on the basic degrees of freedom.

The DKS problem then becomes somewhat more philosophical: such a model predicts at late times that there will be an infinite number of observers, which fluctuate into existence briefly and have an experience very different from those in that first era, so it predicts that we are very atypical observers.
The time scale for the appearance of the more typical observers is of order $e^{m L_P 10^{60}} $, where $m$ is the minimum mass necessary to sustain a system that can be called an observer.   This number is so huge that it's essentially unchanged if we think of the unit of time as the Planck time or the age of the universe.  One can argue\cite{nightmare} that no measuring device can last long enough to verify the existence of more than one of these observers.  Indeed, one can change the model to eliminate this behavior in an infinite number of possible ways, without effecting the predictions of the model for the initial era of cosmological expansion.  That is, the ``phenomenological" disaster caused by stable dS space is not amenable to experimental checks, and can be eliminated in many ways that do not effect any possible future experiment that can be done with localized apparatus existing in the causal diamond of any segment of a timelike trajectory originating on the Big Bang.  

I think that the conservative conclusion that one should draw from these remarks is that all statements that one can make about the correct model of dS space are highly conjectural.  My own preference is to work with mathematically well defined models, which fit the theoretical data provided by effective field theory, rather than to make speculations based on the assumption that the eventual model will have a quantum Hilbert space, which closely resembles that of effective field theory in some background space-time.

\end{document}